# A 16-channel fiber array-coupled superconducting single-photon detector array with average system detection efficiency over 60% at telecom wavelength

**WEI-JUN ZHANG,**[1,2,3,†] **GUANG-ZHAO XU,**[1,2,3] **LI-XING YOU,**[1,2,3,*] **CHENG-JUN ZHANG,**[1,2,3] **HAO HUANG,**[1,3] **XIN OU,**[1,3] **XING-QU SUN,**[1,2,3] **JIA-MIN XIONG,**[1,2,3] **HAO LI,**[1,2,3] **ZHEN WANG,**[1,2,3] **AND XIAO-MING XIE**[1,2,3]

[1]State Key Lab of Functional Materials for Informatics, Shanghai Institute of Microsystem and Information Technology (SIMIT), Chinese Academy of Sciences (CAS), 865 Changning Rd., Shanghai, 200050, P. R. China.
[2]CAS Center for Excellence in Superconducting Electronics (CENSE), 865 Changning Rd., Shanghai, 200050, P. R. China.
[3]Center of Materials Science and Optoelectronics Engineering, University of Chinese Academy of Sciences, 19 A Yuquan Rd, Shi-jingshan District, Beijing 100049, P. R. China
*Corresponding author: [†]zhangweijun@mail.sim.ac.cn, *lxyou@mail.sim.ac.cn.

**We report a compact, scalable, and high-performance superconducting nanowire single-photon detector (SNSPD) array by using a multichannel optical fiber array-coupled configuration. For single pixels with an active area of 18 μm in diameter and illuminated at the telecom wavelength of 1550 nm, we achieved a pixel yield of 13/16 on one chip, an average system detection efficiency of 69% at a dark count rate of 160 cps, a minimum timing jitter of 74 ps, and a maximum count rate of ~40 Mcps. The optical crosstalk coefficient between adjacent channels is better than −60 dB. The performance of the fiber array-coupled detectors is comparable with a standalone detector coupled to a single fiber. Our method is promising for the development of scalable, high-performance, and high-yield SNSPDs.**

Superconducting nanowire single-photon detectors (SNSPDs) [1] have shown excellent performance in visible to infrared (IR) single-photon detection. In the past few years, the key figure merits of SNSPDs have considerably improved, e.g., over 90% system detection efficiency (SDE) at the telecom wavelength of 1550 nm [2-4], sub 1 cps dark count rate (DCR) [5], sub 10 ps timing jitter (TJ) [6], and over 1.5 Gcps maximum count rate (MCR) [7]. SNSPDs have played a significant role in quantum information processing [8, 9], LIDAR applications [10], and deep space optical communication [11]. Recently, high-performance multichannel SNSPDs are needed in emerging applications, especially in multiphoton boson sampling experiments [9] (requiring 100 detectors), large-scale quantum communication networks [8], large-scale integrated optics [12], and single-photon ranging and imaging [13]. However, the current SNSPD usually couples with a single fiber by using self-alignment packaging [4, 14] or metallic housing modules [3, 15]. When a large number of detectors are required, the volume of the entire system becomes very large [9]. Meanwhile, owing to the low detector yield, screening detectors become a time-consuming and laborious task.

Previously, self-aligned four-channel SNSPDs [16] based on a backside Si micromachining process have been demonstrated with an SDE of ~16% and a low optical crosstalk at ~1 Mphoton/s photon flux illumination. However, the high SDE of such detectors may be hindered by the reflection and interference at the multiple interfaces. Additionally, this method still requires individual optical coupling of the fiber to the detector, which is impractical for a channel number of >100. An alternative is front-side coupling the detector array with a multichannel optical fiber array. The fiber array currently represents the most compact and high-density (up to 128 channels) packaged means for multiple fibers with a lateral positioning error of <1 μm [17] by gluing in a V-groove substrate. Previous studies have been reported on fiber array-coupled cryogenic detectors, e.g., a two-pixel TES array [18] and an eight-channel SNSPD chip prototype (fiber array glued on the chip) [19]. However, both of the reported SDEs are quite low (<5%) and thus are less attractive for practical applications. The main challenge here is realizing high-accuracy optical coupling between the detector array and fiber array, especially for detectors operating at cryogenic temperatures. Moreover, the high yield of nanowires is particularly important in research for scalable solutions.

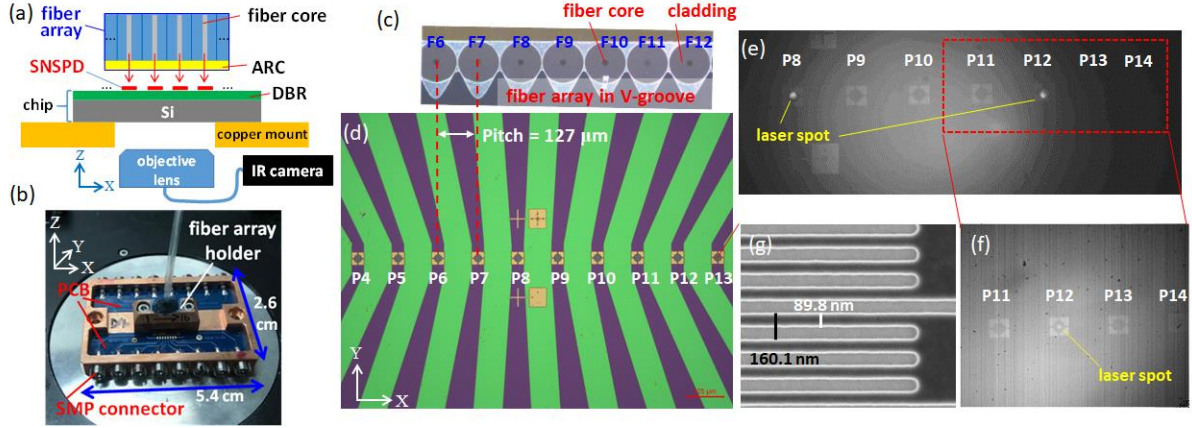

**Fig. 1.** (a) Cross-section sketch of an SNSPD array chip coupled with a fiber array. (b) Photo of the SNSPD array chip block that is packaged with a fiber array holder. (c) Zoomed-in bottom view of a fiber array glued in a V-groove block. The pitch of two adjacent fiber cores is approximately 127 μm (indicated by red dashed lines) and is limited by the V-groove block. (d) Optical image of the SNSPD array. The NbN film, the DBR substrate, and gold alignment mark region are shown in purple, green and golden colors. The active area of a single pixel is surround by a hollow square gold mark. (e) IR camera image acquired at moderate magnification after optical alignment, where the object lens is focused on the sample surface through the Si substrate. Two laser spots emitted from two fibers (F8 and F12) are marked. (f) Zoomed-in IR camera image for the detectors P11 to P14. The inner diameter of a gold mark is ∼24 μm. For pixel P12, the laser spot center almost overlaps the center of the SNSPD. (g) Magnified SEM image of the fabricated parallel nanowires, with a nominal width of 90 nm and a pitch of 160 nm.

In this letter, we demonstrate a compact and scalable package method for a 16-channel (pixel) SNSPD array coupled with a 16-channel fiber array through a metallic housing module. To improve the yield of the samples, we pre-processed NbN thin films by using He ion irradiation [20]. Through accurate optical alignment and packaging, we achieved a pixel yield of 13/16 on one chip (SDE ≥ 50% criterion), an average SDE of 69% at a DCR of 160 cps, a TJ of 74–132 ps, and an MCR of ∼40 Mcps, for single pixels operated at 1550 nm. We also characterized the optical crosstalk coefficient for this new type of array, which is better than −60 dB. Our method is promising for the development of scalable, high-performance, and high-yield SNSPDs.

Figure 1(a) presents a schematic of the optical design of the SNSPD array chip and its front-side coupling setup. To enhance the optical absorption efficiency, the SNSPD array chip comprised a distributed Bragg reflector (DBR) stacked on a Si substrate and the SNSPD (NbN nanowire) array placed on top of the DBR. A commercially available fiber array was aligned to the SNSPD array with matching separation. The distance (air gap) between the fiber array and the SNSPD array chip was preset to ∼30 μm at room temperature to avoid the fiber array from crushing the chip at cryogenic temperature. Thus, the laser spot size emitted from an SMF-28 fiber is estimated to be ≤12 μm in diameter on the surface of the SNSPD array chip. Thus, the active area of a single pixel was designed to be 18 μm in diameter, which was guaranteed an optical coupling efficiency (OCE) over 99% based on calculation. Using a microscope with an IR camera, the optical coupling process was monitored through the Si substrate. A homemade three-axis (X, Y, and θ in the X–Y plane) alignment stage was used to align the fiber array to the detector array.

Figure 1(b) shows a photo of the 16-channel SNSPD array chip block packaged with a fiber array holder. Each pixel is individually biased and read out through wire bonding to the PCB and then connects to SMP connectors. Figure 1(c) shows a magnified bottom-view photo of the fiber array facet. The overall size of the fiber array block is 3.5 × 2.5 mm² [see Fig. S1 of Supplementary Information (SI)]. The chip size is 4.8 × 4.8 mm². Figure 1(d) shows a zoomed-in top view of the SNSPD array chip. For one-to-one matching, both arrays were designed with a pitch of 127 μm (standard pitch). To improve the visibility through the Si substrate to improve the coupling accuracy, reflective align marks comprising a $SiO_2$ dielectric layer and 100 nm-thick gold were fabricated by the standard ultraviolet lithography and lift-off process. To distinguish the fibers and detectors in different positions, we numbered them separately, i.e., from left to right, and element fibers (detectors) were numbered as F1…F16 (P1…P16, respectively). Fibers and detectors with the same number label indicate that there is a one-to-one matching when coupled. The packaged fiber array exhibited no significant aging and element fiber displacement after several thermal cycles.

By aligning two element fibers to two element detectors, we aligned the whole fiber array to the detector array simultaneously. Figure 1(e) and (f) show images captured by the IR camera after the alignment, indicating that a good alignment was achieved. The entire optical coupling process generally takes ∼5–10 min (also the typical time consumption for a single fiber coupling); thus, the fiber array coupling can significantly reduce the assembling time when a large number of detectors are required. Figure 1(g) shows a typical scanning electron microscope image of the fabricated nanowire. To reduce the reset time of the detector, a series two superconducting nanowire avalanche photon detector (series 2-SNAP) configuration [21] was used in this study, which comprised 7.5 nm-thick, 90 nm-wide, and 160 nm-pitch NbN nanowires.

We characterized the SNSPD array in a 16-channel close-cycled G-M cryocooler at a base temperature of 2.2 K. Figure 2(a) presents the bias current ($I_b$)-dependent SDEs for 13 working pixels on one chip (called chip D4), as measured with a photon flux of ∼0.1 Mphoton/s at a 1550 nm wavelength. The uncertainty of the SDE measurement is <3% [3]. Thanks to the use of NbN films pre-processed by He ion irradiation (see Fig. S2 of SI), most of the working pixels show weak saturated SDE plateaus with current dependence, which implied a nearly unity internal efficiency for the nanowires. The suppressed switching current ($I_{sw}$) appears in three pixels (marked with circles in the figure), possibly due to constrictions in the nanowires. Maximum SDEs are distributed in the range of 50%–80% (an average of ∼69%), as determined at a DCR of ∼160 cps. The variations of maximum SDEs in the pixels are mainly attributed to variations of the OCE, except for the constricted pixels. Figure 2(b) demonstrates the maximum SDE of each pixel distributed in space. Here, we determined the yield criterion [marked with a red dashed line in Fig. 2(a) and (b)], where the maximum SDE of the pixel was ≥50%. Interestingly, the 13 working pixels all fulfilled this criterion; thus, a yield of 81% (13/16) was obtained for chip D4. Besides, the three non-functional pixels were caused by two different reasons, i.e., two pixels were nonconductive at room temperature and one pixel showed a normal $I_{sw}$ at 2.2 K; however, its optical input channel was broken. We further measured the TJ and MCR of the pixels (see Fig. S5 of SI). Of the 13 pixels, 10 demonstrated a jitter value that was <90 ps. The jitter among the pixels was 74–132 ps, as measured at $I_b$ of 17.5–13.3 μA. A MCR of ∼40 Mcps for a single pixel is determined. In addition to separately operate the pixels, we also can combine them as a high-speed standalone detector by sharing one optical input through 1 × 16 fiber-based splitters and one electrical signal output through a 16-way RF power combiner [7]. Then, if all 16 pixels are functional, the expected MCR of this array can achieve 640 Mcps.

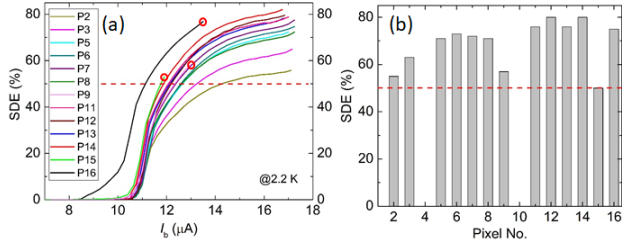

**Fig. 2.** (a) SDE vs. $I_b$ for 13 working pixels of SNSPD array illuminated at 1550 nm and operated at 2.2 K. Red circles mark the pixels with suppressed $I_{sw}$. (b) Maximum SDE vs. the pixel number, showing the spatial distribution of the SDE across the array.

The abovementioned results demonstrate the advantages of integrating the detectors on one chip over the standalone detectors [3]. Because the pixels are distributed in a small area, it is easy to obtain a better uniformity in film thickness and processing, which is good for improving the sample yield (see Fig. S6 of SI). In addition, when the sample uniformity is high enough, the performance of the entire chip can be preliminarily judged by measuring only one of the pixels, which would considerably reduce the screening time of the samples.

Low crosstalk is a key merit of the multipixel array. When pixels of our array are used to detect signals from different optical channels, there is a possibility of optical crosstalk due to the use of a small fiber array pitch. In this situation, the noise from the optical crosstalk detected by one pixel is originated from the reflected photons emitted by the adjacent optical fibers because of the non-flat surface of nanowires and the wave characteristics of light. Multiple reflections appear between the surfaces of the chip and the fiber array facet, as shown in Fig. 3(a). In fact, the DBR substrate covered by the NbN film is a strong IR absorber (i.e., a measured reflectance of <3%@1550 nm [3]) because of the formation of the optical cavity. Thus, the intensity of multiple reflections would expect to decay fast (i.e., low optical crosstalk). We characterized the optical crosstalk of a single optical channel by monitoring the count rate (CR) of one pixel (called $P_m$) with no light input to its own coupled fiber while increasing the input photon rate (IPR) of an adjacent fiber (called $F_{in}$). When the IPR = 0, we experimentally measured the DCR of the $P_m$ (here, P12), which is shown by black dots in Fig. 3(b). An inflection point in DCR vs. $I_b$ appears at ∼16.4 μA (labeled as $I_{DCR-BG}$ in Fig. 3(b)), below which the DCR is dominated by blackbody photons [5]. Thus, a typical background DCR of a single pixel is ∼160 cps at $I_{DCR-BG}$, which is comparable with the one coupled with a single fiber [4]. Then, we increased the IPR of $F_{in}$ (here, F13) from 10 M to 1000 Mphoton/s. The CR increased gradually as the IPR increased. Thus, we define an error CR (ECR) due to optical crosstalk as ECR = $CR_{IPR}$ − $CR_0$. Figure 3(c) shows a plot of the ECR vs. $I_b$ at varied IPRs. Clearly, the ECR from the $F_{in}$ is proportional to the IPR (see linear fits in Fig. S7 of SI). For an IPR of 100 Mphoton/s, the ECR gradually increases from 25 cps at $I_{av}$ = 11.0 μA to 72 cps at $I_{DCR-BG}$ = 16.4 μA. Then, by normalizing the ECR to the corresponding IPR, we estimated the optical crosstalk coefficient of an adjacent optical channel, which produced a value better than −60 dB (<7 × 10$^{-7}$).

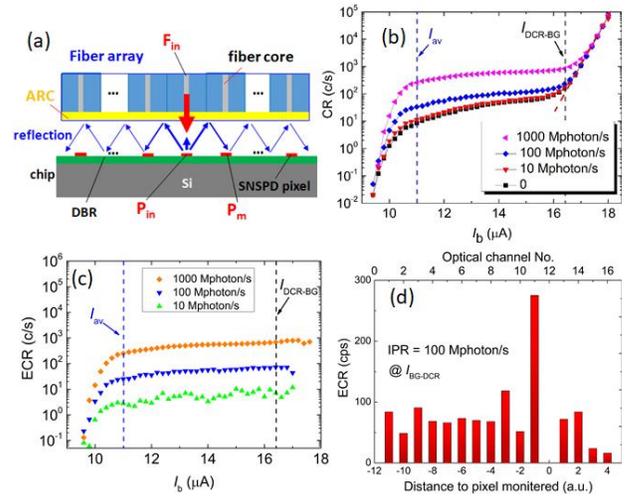

**Fig. 3.** (a) Schematic of reflection-induced optical crosstalk and the principle of the optical crosstalk measurement. (b) CR of the pixel P12 vs. $I_b$ recorded at different IPRs of an adjacent fiber, $F_{in}$. The IPR varies from 0 to 1000 Mphoton/s. The avalanche current ($I_{av}$) of the series 2-SNAP is indicated by the blue arrow. The red dashed line indicates the trend of intrinsic DCR for the high $I_b$ region. (c) ECR vs. $I_b$ recorded at different IPR. (d) ECR vs. the normalized distance from the illuminated fiber to the monitored detector. The ECR was determined at $I_{BG-DCR}$ under an IPR of 100 Mphoton/s. Top tip labels show the optical channel number.

Next, we continuously evaluated the optical crosstalk of $P_m$ (P12) caused by the remaining 13 optical channels by altering the $F_{in}$ channels one by one. To make the optical crosstalk effect significant, an IPR of 100 Mphoton/s is incident to the $F_{in}$, which is two orders of magnitude higher than the ones used under typical working conditions [9]. By doing this, we could also explore the influence of the distance change of the $F_i$ (light source) on the crosstalk of the $P_m$. Figure 3(d) shows the results, where the ECRs of the $P_m$ are plotted as a function of normalized distance (*D*). The *D* is the horizontal distance from the $F_{in}$ to the $P_m$ and is normalized by the pitch of the fiber array. The ECR caused by the $F_{in}$ at *D* = −1 is the highest, reaching a value of 275 cps. This may be due to more direct reflection photons incident to the $P_m$, possibly caused by the fiber array facet not being fully parallel to substrate. We also noticed that, as *D* increased from 0 to −12, the ECR did not show the expected decreasing trend. This result implied that photons were diffusely scattered between the two surfaces of the chip and the fiber array, and a small change in distance (compared with the size of the fiber array facet) had little effect on the overall result. By replacing the fiber array with a single lens fiber, we studied the crosstalk under a single fiber coupling (see Fig. S8 of SI), where the ECR shows a decreasing trend as the distance increases. This result implies that a small reflection cross-section would help to reduce the crosstalk. Simulations (e.g., spatial modeling [22]) are necessary to understand the effects of the crosstalk and provide insight into methods for future reduction.

To approximately estimate the cumulative optical crosstalk of $P_m$ caused by the entire fiber array (corresponding to all 15 adjacent optical channels input at the same time), we added up the ECRs of the $P_m$ shown in Fig. 3(d) as the cumulative ECR, which was ~1208 cps (i.e., an average ECR of 80.5 cps per channel). With the cumulative IPR of 1500 Mphoton/s, a cumulative optical crosstalk coefficient on $P_m$ was obtained that was better than −60 dB (<8.0 × $10^{-7}$). Not surprisingly, the cumulative crosstalk on $P_m$ was nearly equal to the one for a single optical channel because the ECR is weakly distance-dependent and mainly determined by the IPR. Note that an optical crosstalk coefficient of −60 dB is negligible in most applications.

Finally, we discuss future studies that need to be done, such as improving the stability of the package block to withstand multiple thermal cycles (see Fig. S9 of SI) and improving the OCE by either increasing the active area of the pixels or reducing the laser spot size. Further, we may expand the dimension of the array (to 2D), and explore possible ways to reduce the complexity of the readout circuit by using the spatial resolution ability of the fiber array. The biggest issue for the array expansion is to install a large number of cryogenic readout cables in a cryostat; however, the cryostat has a limited cooling power. This problem can be solved by using low-cost, low heat-leakage flexible microstrip circuits [23]. Thus, we believe further scaling of the array is promising.

In conclusion, we have demonstrated a compact, scalable, and practical method for an SNSPD array coupled with a fiber array. Owing to the use of NbN films pre-processed by He ion irradiation, we obtained a 16-pixel SNSPD array with high pixel yield, and high detection performance. The optical crosstalk coefficient among neighboring channels was found to be better than −60 dB, which was caused by the diffuse reflection between the surface of the chip and the fiber array facet. We believe these results have implications for multiphoton boson sampling experiments, large-scale quantum communication networks, and multi-beam photon-counting LIDAR and imaging.

**Funding.** National Natural Science Foundation of China (61971409), National Key R&D Program of China (2017YFA0304000), and Science and Technology Commission of Shanghai Municipality (18511110202). W. J. Zhang is supported by the Youth Innovation Promotion Association, CAS (2019238).

**Disclosures**. The authors declare no conflicts of interest.

See Supplement 1 for supporting content.

**REFERENCES**

1. G. N. Gol'tsman, O. Okunev, G. Chulkova, A. Lipatov, A. Semenov, K. Smirnov, B. Voronov, A. Dzardanov, C. Williams, and R. Sobolewski, Appl. Phys. Lett. **79**, 705 (2001).
2. F. Marsili, V. B. Verma, J. A. Stern, S. Harrington, A. E. Lita, T. Gerrits, I. Vayshenker, B. Baek, M. D. Shaw, R. P. Mirin, and S. W. Nam, Nat. Photon. **7**, 210 (2013).
3. W. J. Zhang, L. X. You, H. Li, J. Huang, C. L. Lv, L. Zhang, X. Y. Liu, J. J. Wu, Z. Wang, and X. M. Xie, Sci. China Phys. Mech. Astron. **60**, 120314 (2017).
4. I. E. Zadeh, J. W. N. Los, R. B. M. Gourgues, V. Steinmetz, G. Bulgarini, S. M. Dobrovolskiy, V. Zwiller, and S. N. Dorenbos, APL Photon. **2**, 111301 (2017).
5. H. Shibata, K. Shimizu, H. Takesue, and Y. Tokura, Opt. Lett. **40**, 3428 (2015).
6. J. P. Allmaras, A. G. Kozorezov, B. A. Korzh, K. K. Berggren, and M. D. Shaw, Physical Review Applied **11**, 034062 (2019).
7. W. J. Zhang, J. Huang, C. J. Zhang, L. X. You, C. L. Lv, L. Zhang, H. Li, Z. Wang, and X. M. Xie, IEEE Trans. Appl. Supercond. **29**, 1 (2019).
8. Q. C. Sun, Y. F. Jiang, Y. L. Mao, L. X. You, W. Zhang, W. J. Zhang, X. Jiang, T. Y. Chen, H. Li, Y. D. Huang, X. F. Chen, Z. Wang, J. Y. Fan, Q. Zhang, and J. W. Pan, Optica **4**, 1214 (2017).
9. H.-S. Zhong, H. Wang, Y.-H. Deng, M.-C. Chen, L.-C. Peng, Y.-H. Luo, J. Qin, D. Wu, X. Ding, Y. Hu, P. Hu, X.-Y. Yang, W.-J. Zhang, H. Li, Y. Li, X. Jiang, L. Gan, G. Yang, L. You, Z. Wang, L. Li, N.-L. Liu, C.-Y. Lu, and J.-W. Pan, Science **370**, 1460 (2020).
10. L. Xue, Z. Li, L. Zhang, D. Zhai, Y. Li, S. Zhang, M. Li, L. Kang, J. Chen, P. Wu, and Y. Xiong, Opt. Lett. **41**, 3848 (2016).
11. M. E. Grein, A. J. Kerman, E. A. Dauler, M. M. Willis, B. Romkey, R. J. Molnar, B. S. Robinson, D. V. Murphy, and D. M. Boroson, SPIE Sens. Technol. + Appl. **9492**, 949208 (2015).
12. J. Wang, F. Sciarrino, A. Laing, and M. G. Thompson, Nat. Photon. **14**, 273 (2020).
13. A. McCarthy, N. J. Krichel, N. R. Gemmell, X. Ren, M. G. Tanner, S. N. Dorenbos, V. Zwiller, R. H. Hadfield, and G. S. Buller, Opt. Express **21**, 8904 (2013).
14. A. J. Miller, A. E. Lita, B. Calkins, I. Vayshenker, S. M. Gruber, and S. Nam, Opt. Express **19** (2011).
15. S. Miki, M. Takeda, M. Fujiwara, M. Sasaki, and Z. Wang, Opt. Express **17** (2009).
16. R. Cheng, X. Guo, X. Ma, L. Fan, K. Y. Fong, M. Poot, and H. X. Tang, Opt. Express **24**, 27070 (2016).
17. J. H. C. V. Zantvoort, S. G. L. Plukker, E. C. A. Dekkers, G. D. Petkov, G. D. Khoe, A. M. J. Koonen, and H. Waardt, IEEE J. Sel. Top. Quant. Electron. **12**, 931 (2006).
18. L. Lolli, E. Taralli, C. Portesi, D. Alberto, M. Rajteri, and E. Monticone, IEEE Trans. Appl. Supercond. **21**, 215 (2011).
19. F. Bellei, D. Zhu, H. Choi, L. Archer, J. Mower, D. R. Englund, and K. Berggren, CLEO: QELS_Fund. Sci., FTu4C.3 (2016).
20. W. Zhang, Q. Jia, L. You, X. Ou, H. Huang, L. Zhang, H. Li, Z. Wang, and X. Xie, Phys. Rev. Appl. **12**, 044040 (2019).
21. F. Najafi, A. Dane, F. Bellei, Q. Zhao, K. Sunter, A. N. McCaughan, and K. K. Berggren, IEEE J. Sel. Top. Quant. Electron. **21**, 1 (2015).
22. B. Piccione, X. Jiang, and M. A. Itzler, Opt. Express **24**, 10635 (2016).
23. A. I. Harris, M. Sieth, J. M. Lau, S. E. Church, L. A. Samoska, and K. Cleary, Rev Sci Instrum **83**, 086105 (2012).

# Supplementary Information:

# A 16-channel fiber array-coupled superconducting single-photon detector array with average system detection efficiency over 60% at telecom wavelength

WEI-JUN ZHANG,[1,2,3,†] GUANG-ZHAO XU,[1,2,3] LI-XING YOU,[1,2,3,*] CHENG-JUN ZHANG,[1,2,3] HAO HUANG,[1,3] XIN OU,[1,3] XING-QU SUN,[1,2,3] JIA-MIN XIONG,[1,2,3] HAO LI,[1,2,3] ZHEN WANG,[1,2,3] AND XIAO-MING XIE[1,2,3]

[1]State Key Lab of Functional Materials for Informatics, Shanghai Institute of Microsystem and Information Technology (SIMIT), Chinese Academy of Sciences (CAS), 865 Changning Rd., Shanghai, 200050, P. R. China.
[2]CAS Center for Excellence in Superconducting Electronics (CENSE), 865 Changning Rd., Shanghai, 200050, P. R. China.
[3]Center of Materials Science and Optoelectronics Engineering, University of Chinese Academy of Sciences, 19 A Yuquan Rd, Shi-jingshan District, Beijing 100049, P. R. China
*Corresponding author: †zhangweijun@mail.sim.ac.cn, *lxyou@mail.sim.ac.cn.

1. Fiber array

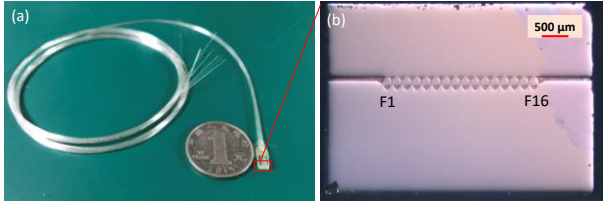

**Fig. S1.** (a) Photo of an unpackaged 16-channel fiber array, compared with a 1-yuan coin. (b) Zoom-in microscope photo of the facet of the fiber array.

Figure S1(a) shows photos of an unpackaged fiber array with 16-bundle bare fiber pigtails. An anti-reflection coating (ARC) layer was deposited on the facet with a target wavelength of 1550 nm and a 3 dB bandwidth of ±100 nm to reduce the interface reflection (fiber to air) of the fiber array. The ARC layer appears pink in the optical photo. Figure S1(b) shows a micrograph of the fiber array facet. Sixteen element fibers (numbered as F1…F16) were fixed to the V-groove block using low-expansion glue.

2. Detectors fabricated with/without ion-irradiated NbN films

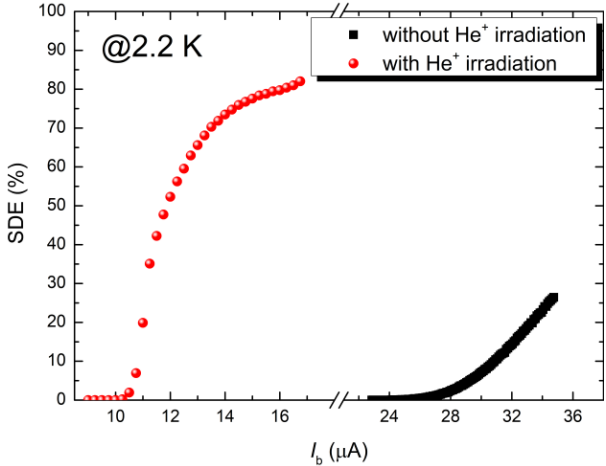

**Fig. S2.** Typical SDE as a function of bias current for the detectors fabricated with (red dots) or without (black squares) ion-irradiated NbN films, as measured in a 2.2 K G-M cryocooler.

In this study, NbN films were deposited in the same batch with a nominal thickness of 7.5 nm by monitoring the deposition time. We conducted He ion irradiation with a fluence of $5 \times 10^{16}$ ions cm$^{-2}$ at an ion energy of 20 keV using an 8 inch ion implanter [1]. The $T_c$ values of the NbN films with or without He ion-irradiated NbN films were 7.25 and 8.33 K, respectively. The resistivities (determined at 12 K) of the films with or without He ion-irradiated NbN films were ~461 and 380 μΩ·cm, respectively. The suppression ratio of the $I_{sw}$ due to irradiation was ~2. Figure S2 shows a comparison of the SDE($I_b$) curves for detectors fabricated with or without He ion-irradiated NbN films. It was found that the SDE($I_b$) curve of the detector without irradiation was unsaturated, and the maximum SDE was ~26%. By contrast, a weak saturation SDE plateau appeared for the detector fabricated with the irradiated NbN films because of the reduction in $T_c$ and the increase in resistivity in nanowires can help to create a larger hotspot [2]. When coupled to the fiber array (or single lens fiber), the maximum SDE of the irradiated detector was ~81% (~90%, not shown). The He ion irradiation process was well reproduced and controllable; thus, it can be used as a standard process for SNSPD fabrication.

3. Wavelength dependence of SDE at around 1550 nm

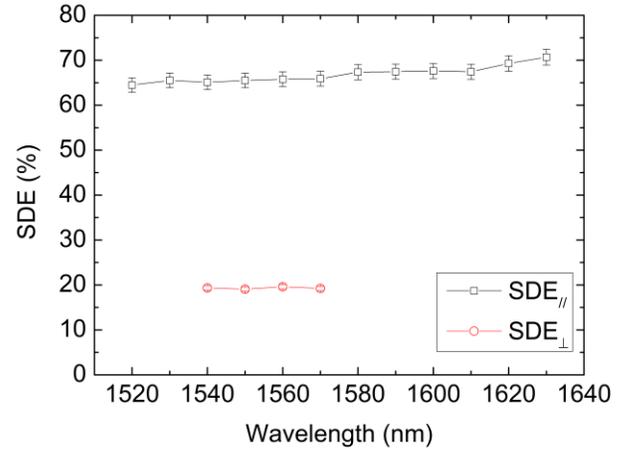

**Fig. S3.** Wavelength-dependent SDE for parallel (||) and perpendicular (⊥) polarized light in the range of 1520–1630 nm.

Figure 3 shows the SDE of a single pixel as a function of wavelength in the range of 1520–1630 nm, as measured at two orthogonal polarization states of incident light. The SDE was weakly wavelength dependent because of the broadband design of the optical cavity and the ARC of the fiber array facet. The maximum SDE (SDE$_{||}$) and minimum SDE (SDE$_\perp$) were ~65% and 19% at 1550 nm, respectively, which produced a polarization extinction ratio of ~3.4.

4. Temperature dependence of the SDE

Figure S4 shows two typical SDE($I_b$) curves for a series 2-SNAP pixel (P8 of chip D1), as measured at two different temperatures, showing the influence of the operating temperature. As the temperature decreased, the rising edge of the SDE($I_b$) curve shifted to a higher current region, and the saturated SDE plateau became broader. Specifically, the $I_{av}$ shifted from 11.0 μA at 2.2 K to the higher value of 12.2 μA at 1.4 K. This was because $I_{av}$ is proportional to $I_{sw}$, which is temperature dependent. The width of the saturation SDE plateau reached ~3 μA at 1.4 K. For the present chip array, $I_{av}$ was ~0.65$I_{sw}$. Obviously, lowering the operating temperature of the detector would help to improve the yield of the detectors. Moreover, we determined the saturation current, $I_{sat}$, above which the detector began to exhibit a saturated SDE plateau [or internal efficiency (IE) ~100%]. The sigmoidal fit for the 2.2 K data implied that the IE of the pixel reached saturation when the $I_b$ was ≥15.0 μA.

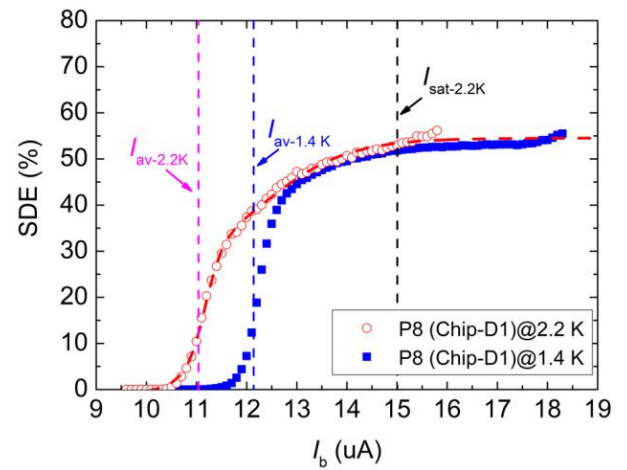

**Fig. S4.** SDE vs. $I_b$ for the pixel P8 on chip D1, measured at 2.2 K (red dots) and 1.4 K (blue squares) in a 1-K close-cycled cryocooler. The magenta and black arrows indicate the $I_{av}$ and $I_{sat}$ determined at 2.2 K. The blue arrow indicates the $I_{av}$ at 1.4 K. Red dashed line is a sigmoidal fit.

5. TJ performance and maximum CR of the single pixels

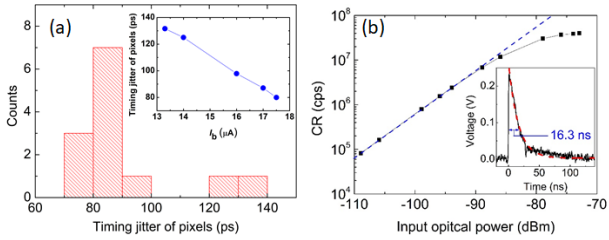

**Fig. S5.** (a) Histogram of the timing jitter for the 13 working pixels. The inset graph indicates the optimal jitter of five typical pixels vs. $I_b$. (b) Count rate vs. input optical power. The inset graph indicates typical photon-response pulse of a single pixel. Red-dashed line shows the exponential fit for the falling edge of the pulse.

Figure S5(a) shows the histogram of the TJ measurement for the 13 working pixels. We measured the TJ using a TCSPC 150 module and a synchronous fs-laser source [3]. The average TJ of the 13 pixels is ~85.3 ps, distributed in a range of 70–140 ps. Most of the pixels (10/13) demonstrated a jitter value that was <90 ps. The inset of Fig. S5(b) shows the optimal time jitters for five typical pixels as a function of $I_b$. The minimum (maximum) jitter among the pixels was ~74 (132) ps, as measured at $I_b$ of 17.5 (13.3) μA, respectively. Figure S5(b) shows the typical input optical power-dependent CR for a single pixel. When CR ≥6.8 Mcps, the curve began to deviate from linearity; upon further increasing the input power, the CR reached a saturated value of ~40 Mcps, which was determined to be the MCR. Because of the limit of the readout circuit, the MCR was less than the calculation deduced from the fitted exponential (1/e) decay time of 16.3 ns for the photon-response pulse [see Inset of Fig. S5(d)].

6. Switching current distribution for chips in one wafer

Figure S6 shows the pre-screening result of our chips fabricated in one wafer by measuring the $I_{sw}$ of the single pixels at 2.2 K. The chips were designed with the same width and pitch of the nanowires using a seris-2SNAP structure. Thus, we assumed the normal pixels share the same $I_{sat}$ ~15.0 at 2.2 K under the 1550 nm photon illumination, as determined in Fig. S4. The empirical assumption holds when the differences of $I_{sw}$ of the pixels are small. With the current criterion, we can roughly determine which chip is usable. For example, chip D1 has 14 of 16 pixels with $I_{sw}$ > $I_{sat}$. For

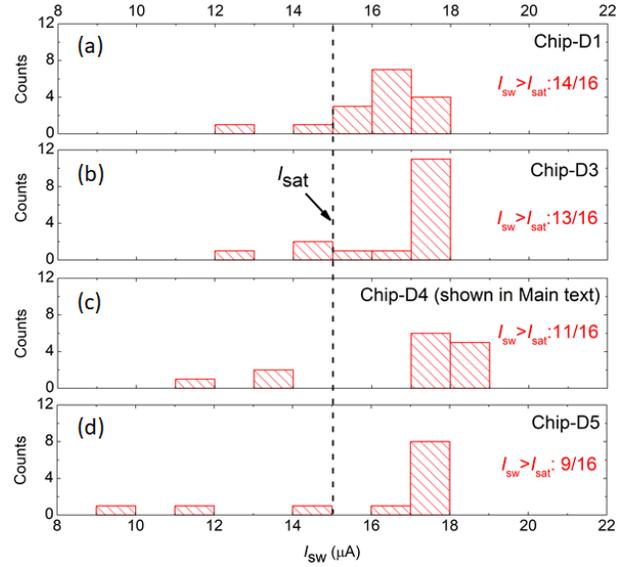

**Fig. S6.** Histogram of the $I_{sw}$ for pixels of four different chips (a)–(d) in one wafer, measured at 2.2 K. The black arrow indicates the $I_{sat}$ determined at 2.2K for a typical pixel.

Chip D4, the $I_{sw}$ of most pixels (11/16) are distributed in the small range of 18 ± 1 μA. Thus, Chip D4 was selected for further characterization. As shown in Fig. 2 of the main text, 13 of 16 pixels were working, and their performances were characterized. For the 4 chips shown in the Figure S6(a-d), we observed 47 pixels of the measured 64 pixels with $I_{sw}$ > $I_{sat}$, giving an overall ratio of 73.4%.

7. Linear fitting of ECR as a function of IPR

Figure S7 shows the ECR determined at two different currents as a function of the IPR of adjacent fiber. Dashed lines represent the linear fits for the experimental data. It can be observed that, the data fits the linear relation quite well (the $R^2$ values for the fits ~1), implying that the detector linearly responses to the scattered photons, and the ECR is dominated by the IPR.

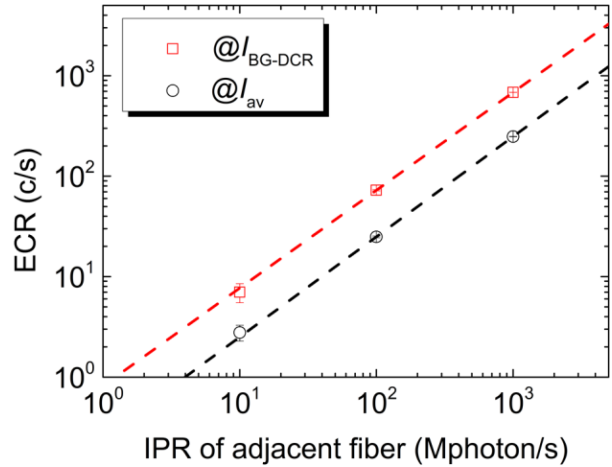

**Fig. S7.** ECR vs. IPR of adjacent fiber when the detector is biased at the current of $I_{BG-DCR}$ and $I_{av}$, respectively. Dashed lines represent linear fits for the measured data (open scatters).

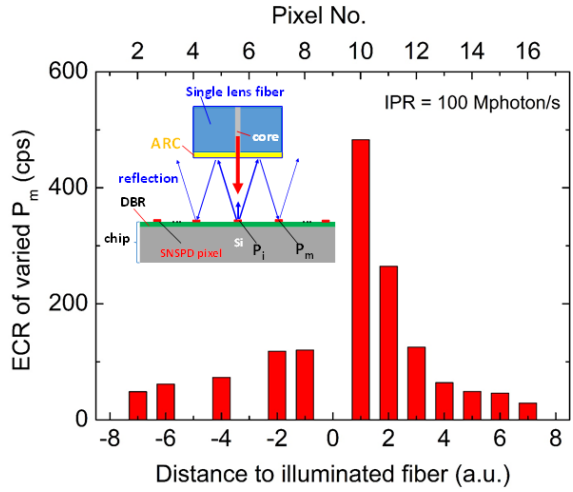

**Fig. S8.** ECR of varied $P_m$ vs. the normalized distance from monitored detector to illuminated fiber. Inset, a schematic diagram of a single lens fiber coupled SNSPD array and the optical crosstalk of $P_m$ caused by the reflection of photons emitted from the fiber. Top tip labels show the actual pixel number (the position of the $P_m$).

8. Optical crosstalk under a single fiber coupling

To further study the effect of photon reflection on the optical crosstalk, we coupled one pixel (P9) of the detector array to a single lens fiber. The facet size of the GRIN lens fiber was ∼600 μm in diameter, which was much smaller than the facet of the fiber array. Additionally, the facet to the surface of the chip was ∼150 μm in the vertical direction, which was also much larger than the one (∼30 μm) for the fiber array to the chip. In the experiment, we only had one incident fiber; therefore, we changed the monitoring pixel ($P_m$) one by one to investigate the influence of the lens fiber on the optical crosstalk of the detectors at different horizontal distances ($D$).

Figure S8 shows the results of the measurement. The distance was normalized by the pitch of the fiber array. We observed a slightly symmetrical ECR distribution. The air gap between the fiber and chip had little effect, i.e., the reflection did not reduce as the gap increased. The highest ECR appeared at P10, reaching a value of 500 cps, which was higher than the peak value (275 cps) shown in Fig. 3(d). We speculate that this may be because the facet of the lens fiber was more parallel to the surface of the chip. We found that as the horizontal distance (($|D|>2$) from $P_m$ to the fiber increased (became larger than the fiber facet), the ECR reduced quickly to ∼50 cps, indicating that the reflection cross-section of the fiber facet contributed to the ECR. Notably, the $P_m$ at positions ($D$ = −8, −5, and −3) were not working.

9. Thermal cycles

The thermal cycle stability of mechanical packaging is significant for long-term practical applications of the detector array. Figure S9(a) shows a photo of the chip package block mounted on the 2-K stage of the G-M cryocooler before cooling the system down. Owing to the compact design of the packaging block, the current sample stage can support two such blocks, i.e., if there is enough cryogenic coaxial cable installed, a 32-channel detector system is achievable.

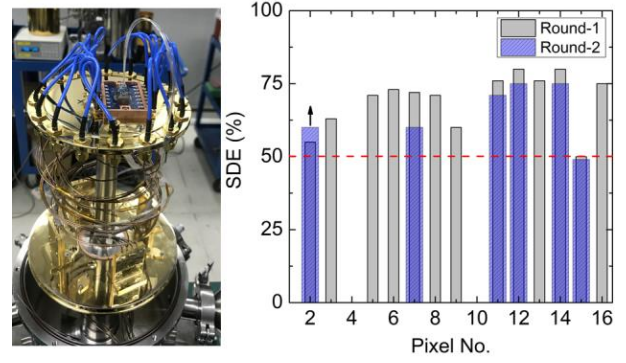

**Fig. S9.** (a) Photograph of the chip package block mounted on the 2-K stage of a G-M cryocooler. (b) Spatial distribution of SDEs for the chip D4 as recorded by two consecutive thermal cycle measurements. The upward arrow marks the pixel with increased SDE undergoing a thermal cycle.

To test the stability of our detector for undergoing thermal cycles, we measured and compared the spatial SDE distributions of the pixels between two consecutive thermal cycles. Figure S9(b) shows the results. The data from the first round has been shown in Fig. 2 of the main text. In the second measurement, we selected six pixels as reference pixels, which were distributed on both sides of the fiber array, as shown by blue bars in the figure. It was found that only the SDE of one pixel (P2) increased from 55% to 60%. However, the SDEs of the remaining five pixels all decreased with a reduction of 2%–12%. This result illustrates that the packaged fiber holder underwent subtle shifting during the thermal cycles owing to thermal expansion and contraction. Thus, in the future, we will permanently seal the moving parts of the package block with cryogenic cement to reduce the probability of the packaging block shifting.

**REFERENCES**

1. W. Zhang, Q. Jia, L. You, X. Ou, H. Huang, L. Zhang, H. Li, Z. Wang, and X. Xie, Phys. Rev. Appl. 12, 044040 (2019).
2. A. Engel, J. J. Renema, K. Il'in, and A. Semenov, Supercond. Sci. Technol. 28, 114003 (2015).
3. W. J. Zhang, L. X. You, H. Li, J. Huang, C. L. Lv, L. Zhang, X. Y. Liu, J. J. Wu, Z. Wang, and X. M. Xie, Sci. China Phys. Mech. Astron. 60, 120314 (2017).